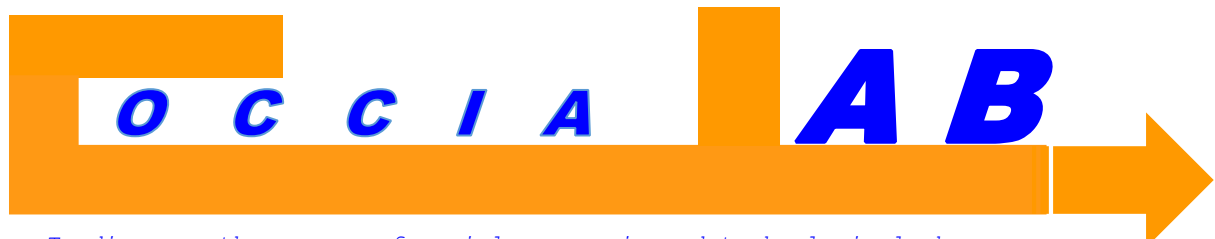

*To discover the causes of social, economic and technological change*

*CocciaLAB Working Paper 2018 – No. 29*

# Measurement of the evolution of technology: A new perspective

Mario COCCIA

CNR -- NATIONAL RESEARCH COUNCIL OF ITALY

&

ARIZONA STATE UNIVERSITY

# Measurement of the evolution of technology: A new perspective

*Mario Coccia*
CNR -- National Research Council of Italy &
Arizona State University

Current Address: C<small>occia</small>LAB at CNR -- National Research Council of Italy
Via Real Collegio, 30, 10024-Moncalieri (Torino), Italy

*E*-mail: mario.coccia@cnr.it

Mario Coccia ORCID: http://orcid.org/0000-0003-1957-6731

**Abstract.** A fundamental problem in technological studies is how to measure the evolution of technology. The literature has suggested several approaches to measuring the level of technology (or state-of-the-art) and changes in technology. However, the measurement of technological advances and technological evolution is often a complex and elusive topic in science. The study here starts by establishing a conceptual framework of technological evolution based on the theory of technological parasitism, in broad analogy with biology. Then, the measurement of the evolution of technology is modelled in terms of morphological changes within complex systems considering the interaction between a host technology and its subsystems of technology. The coefficient of evolutionary growth of the model here indicates the grade and type of the evolutionary route of a technology. This coefficient is quantified in real instances using historical data of farm tractor, freight locomotive and electricity generation technology in steam-powered plants and internal-combustion plants. Overall, then, it seems that the approach here is appropriate in grasping the typology of evolution of complex systems of technology and in predicting which technologies are likeliest to evolve rapidly.

**Introduction**

Technical change has a vital role for economic growth of nations and many studies endeavor to explain its sources, dynamics, technology transfer, effects and evolutionary paths in society (Coccia, 2010a, 2015, 2017d)[1]. Patterns of technological innovation have been analyzed using many analogies with biological phenomena (Basalla, 1988; Farrell, 1993; Nelson and Winter, 1982; Solé et al., 2013; Sahal, 1981; Wagner, 2011; Ziman, 2000). Wagner and Rosen (2014) argue that the application of Darwinian and evolutionary biological thinking to different research fields has reduced the distance between life sciences and social sciences generating new approaches, such as the evolutionary theory of economic change (Nelson and Winter, 1982; cf., Dosi, 1988). In the research field of technical change, the measurement of technological advances is a central and enduring research theme to explain the dynamics of the evolution of technology and technological change (Coccia, 2005, 2005a). Scholars in these research topics endeavor of measuring technological advances, the level of technological development and changes in technology with different approaches (Coccia, 2005; Dodson, 1985; Faust, 1990; Fisher and Pry, 1971; Farrell, 1993; Knight, 1985; Martino, 1985; Sahal, 1981; Wang et al., 2016). However, a technometrics that measures and assesses the comprehensive evolution of technology as a complex system of technologies is unknown.

This study confronts this problem by proposing a theory of measurement of the evolution of technology, based on interaction between technologies that may be useful for bringing a new perspective to explain and generalize, whenever possible, the long-run coevolution between technologies in complex systems. In order to position this paper in existing frameworks, the study here starts by establishing general notions of the theory of measurement and a theoretical framework of different approaches for measuring technological advances from engineering, economics and related disciplines. Moreover, in broad analogy with biology, a conceptual framework of the technological evolution is suggested: theory of technological parasitism (Coccia and Watts, 2018). Then, the technometrics of the evolution of technology is modeled in simple way in terms of morphological changes between a host technology and its technological subsystems. The coefficient of evolutionary growth of the proposed model is quantified in real instances using historical data. Overall, then, it seems that the technometrics here is appropriate in grasping the typology of the evolution of technology. This approach also provides fruitful information to predict which technologies are likeliest to evolve rapidly and lays a foundation for the development of more sophisticated concepts to measure and explain the general properties of the evolution of new technology.

---

[1] *See* Calabrese et al., 2002; 2005; Calcatelli et al., 2003; Cavallo et al., 2014, 2015; Chagpar and Coccia, 2012; Coccia, 1999, 2001, 2003, 2004, 2005b, 2005c, 2006, 2008, 2009, 2010, 2010a, 2010b, 2010c, 2011, 2012, 2012a, 2012b, 2013, 2013a, 2014, 2014a, 2014b, 2014c, 2014d, 2015, 2015a, 2015b, 2016, 2017b, 2017c, 2017d, 2017e, 2017f, 2017g, 2017h, 2017i, 2018a, 2018b, 2018c; Coccia and Cadario, 2014; Coccia et al., 2010, 2015; Coccia and Finardi, 2012; Coccia and Rolfo, 2009, 2010, 2013; Coccia and Wang, 2015, 2016; Ferrari et al., 2013.

**Measurement theory: general concepts**

Measurement assigns mathematical characteristics to conceptual entities. Caws (1959, pp. 4-5) argues that: "the result of a measurement is a proposition expressing a relation between a number and an object to which it is assigned . . . . the setting in order of a class of events with respect to its exhibition of a particular property, and this entails the discovery of an ordered class the elements of which can be put in one-to-one correlation with the events in question". Stevens (1959, p. 19) claims that the measurement is: "the assignment of numeral[2] to objects or events". Moreover, Russel (1937, p. 176) posits that: "Measurement demands some one-one relation between the numbers and magnitudes in question – a relation which may be direct or indirect, important or trivial, according the circumstances". In this research field, Campbell (1928) argues that direct measurement is possible only when the "axioms of additivity" can be shown to be isomorphic with the manipulations performed on objects: length and weight are measurable in this way, whereas other magnitudes are measured with indirect measurement by means of numerical laws. In short, the definition of measurement can be generalized to consider the determination of any kind of relation between properties of objects or events. However in general, measurement is restricted to relations for which one or another property of the real number system might serve as a useful model (Stevens, 1959, p. 24).

---

[2] The term "numeral" according to Stevens (1959, p. 19) refers to an element in a formal model, not to a particular mark on a particular piece of paper.

Wilks (1961) states that measurement should have some basic requirements:

- Operationally definable process by specifying a set of realizable experimental conditions and a sequence of operations to be made under these conditions, which will yield the measurement.
- Reproducibility of outcomes: repeating the process should yield measurement in "reasonable agreement" with each other; this property of reproducibility of measurement provides reliability.
- Validity and accuracy of the process of measurement: true measurement of the object being measured.

In the theory of measurement, the fact that numerals can be assigned under different rules leads to different kinds of measurement and of scales (nominal, ordinal, interval and ratio; Table 1).

Table 1. Classification of scale of measurement

| Scale | Empirical operations | Mathematical transformations that leave the scale structure invariant ($x$ and $x'$ are numerals) | Examples |
|---|---|---|---|
| Nominal | a) Determination of equality | Permutation group $x'=f(x)$ $f(x)$ means any one-to-one substitution | Numbering the football players |
| Ordinal | a)+ b) Determination of greater or less | Isotonic group $x'=f(x)$ $f(x)$ means any increasing monotonic function | Hardness of bolt Street numbers |
| Interval | a)+b) + c) Determination of the equality of intervals or of differences | Linear group $x'=ax+b$ $a>0$ | Temperature Time |
| Ratio | a)+b)+c)+ d) determination of the equality of ratios | Similarity group $x'=cx$ $c>0$ | Density, Lengths |

*Note*: a fifth kind of scale never put in use is "logarithmic interval scale" based on three empirical operations: determination of equality, of greater and less, of equal ratios, i.e., identify items, order them and set them in relation a/c=b/c=c/d …in fact, *log* a –*log* b =*log* b –*log* c, etc. The mathematical transformation under which the structure of the scale is invariant is *power group*: for any numerical value $x$, we can substitute $x'$, where $x'=kx^n$ ($k=0$; $n>0$).
*Source*: adapted from Stevens (1959, p. 25).

In general, fundamental measurement of a set *A* is always done with respect to an empirical system 𝔄.

In contrast, derived measurement does not depend on an empirical relational system directly but on other numerical assignments, such as the density defined as the ratio of mass and volume. In addition, the central issue for a theory of measurement of derived measurement is the status of the two basic problems of fundamental measurement: the representational and uniqueness problems (Suppes and Zinnes, 1963, pp. 17ff).

In particular, the first is the justification of the assignment of the numbers to objects or phenomena (called the representational theorem); the second is the specification of the degree to which this assignment is unique (The uniqueness theorem; cf., Suppes and Zinnes, 1963; Luce et al. 1963).

Hence, the measurement can be considered the process of mapping empirical properties or relations into a formal model, considering the two fundamental problems just mentioned.

**Theoretical framework of the measurement of technological advances (technometrics)**

Technometrics refers to a theoretical framework for the measurement of technology and technological change with policy implications (Sahal, 1981). The measurement of technological advances has been performed with different approaches. This section presents some of the most important methods of technometrics, without pretending to be comprehensive.

*Hedonic approach to the measurement of technology*

The basic hypothesis of this approach is simply that the utility of a product or service lays in its essential attributes or quality characteristics, such as size, power, comfort, and fuel economy in the case of an automobile. It is postulated that price variation among many different models and varieties of a particular commodity, at any given point in time, is related to differences in their quality characteristics. Once the set of attributes is selected and the form of the functional relationship specified, observed changes in the price of a commodity can be decomposed into a "quality or technological change effect" and a "pure price effect". The assumption of this approach is a positive relationship between market price of a good or service and its quality. In particular, it is assumed that any model of a particular product can be represented by a set of characteristics and by their value; then, the quality of the product is assumed to be a function of the defining characteristics (of their quantitative levels) and of their relative importance:

$$Q_j = f(a_1,...,a_n, X_{1j},...,X_{2j},...,X_{kj}) \qquad (1)$$

where $a_i$ represents the relative importance of the $i$-th characteristics and $X_{ij}$ the qualitative level of the same characteristics in product model $j$. Technological progress or technology change in a given period of time can be defined as the change in quality during a given period of time:

$$TC_j = \frac{\Delta Q_j}{\Delta t} \qquad (2)$$

Both quality and technological progress, as defined in the first and second equation can be calculated at different levels of aggregation, starting from an individual product model up to that of a firm, an industry, a country, etc. The simplest form of functional relationship between quality and product is a linear combination:

$$Q_j = \sum_{i=1}^{n} a_i X_{ij}$$

Using regression techniques, the observed changes in the price of a commodity can be decomposed into a "quality/technological change" effect and "pure price effect" (i.e., residual). It is important to recognize that the hedonic approach rests on the implicit assumption of a competitive market. If the market is imperfect, however, the interpretation of the price-quality relationship becomes somewhat problematic (cf., Saviotti, 1985).

*RAND[3] approach to the measurement of technology*

A technological device has many technical parameters that measure its characteristics and characterize the state of the art (SOA). Many approaches have tried to measure the SOA and advances in SOA. When technology is characterized by multiple parameters, some means of combining these into a single measure of performance is required. Dodson (1985) specifies the state-of-the-art (SOA) surface for any

---

[3] RAND Corporation ("Research ANd Development") is an American nonprofit global policy think tank created in 1948 by Douglas Aircraft Company to offer research and analysis to the United States Armed Forces. It is financed by the U.S. government and private endowment, corporations, universities and private individuals.

given field as a convex surface in an *N*-dimensional space, where *N* is the number of essential characteristics of a technology. He proposes the use of either a planar or an ellipsoidal surface, given by the following equations:

*Planar*                                              *Ellipsoidal*

$$\sum_{i=1}^{n}\left(\frac{x_i}{a_i}\right)=1 \qquad\qquad\qquad \sum_{i=1}^{n}\left(\frac{x_i}{a_i}\right)^2=1$$

Where $x_i$ is the *i*-th technological characteristic and $a_i$ the *i*-th parameter (a constant). The values of the parameters are obtained by fitting a surface to the data points. Dodson (1985) also suggests the use of multiple regression techniques to study technological indicators (e.g., cost of a technology) in terms of functional measures. Conceptually, the SOA can be viewed as a surface in some multidimensional parameter space. Designers are free to move about on that surface in generating alternative designs. The precise point on the trade-off surface, which a designer chooses, depends upon the user's needs. Dodson (1985) mainly utilizes trade-off surfaces in a dimensional space of order 2 (ellipsoids) that should be convex outward, to allow for the fact that a trade-off becomes more and more difficult as it is pushed further (cf., Coccia, 2005a).

Alexander and Nelson (1973) developed an alternative procedure. They used multiple dimensions, like Dodson (1985), but employed hyperplanes instead of ellipses. The major change is that in Dodson's model the surface was limited to ellipsoids of order 2. The method was extended to allow for the use of ellipsoids of any order. Thus the surface to be fitted to the data was given by the equation:

$$\sum_{i=1}^{n}\left(\frac{x_i}{a_i}\right)^n=1$$

Where *N* is the order of the ellipsoid, $x_i$ the value of the *i*-th variable, and $a_i$ the intercept of the ellipsoid on the *i*-th axis. The concept is that at any given state-of-the-art, the designer has the freedom to obtain more than one parameter at the expense of some other parameters. In general, all technologies having the same score fall on the same surface in the *N*-dimensional space of the parameters. In short, surface represents all the possible combinations of factors at a given state-of-the-art. An advance in the state-of-the-art means that more than one parameter can be obtained without sacrificing any of the remaining parameters. This advance in the state-of-the-art means moving to a higher trade-off surface. Since the trade-off surface is the set of all the points that are just barely reachable at the current state-of-the-art, the surface represents a technology frontier in the *N*-dimensional parameter space. Points outside the frontier cannot be reached at the current state-of-the-art, whereas points inside the frontier are below the current state-of-the-art. Martino (1985; 1993) applies these tradeoff surfaces by Alexander and Nelson (1973) for measuring technological advances of clipper ships, propeller-driven aircraft and power transistors. Overall, then, the hedonic and the RAND techniques just mentioned are very similar and differ only in their choice of the dependent variable, which is price in the former and calendar year in the latter.

*Functional and structural measurement of technology*

The technique by Knight (1985) is based on a functional and a structural description of a given technology over time to depict the evolution of technology. The structural model was originated by Burks et al. (1946) that describe the computing system by outlining, in general terms, the pieces of equipment the computer must have, the purpose of the devices, and the way the items interact with one another to perform as a computer. The functional description of a new computer over an earlier one indicates that technological advancement has taken place, but it does not specify the details of the new development. Therefore in order to define the technological advances, it is necessary to use the structural description. Knight (1985) identifies the developments by comparing the structure of new systems with that of earlier computers. The resulting structural changes represent the new technology introduced. The structural elements of a computer have a clearly defined relationship to its power. It is possible to use these elements to obtain measures of performance. For instance, if a computer's circuits suddenly begin operating at twice their pervious speed, the central processing unit would then perform twice as many operations in any given period of time. A regression technique can be used to determine cost as a function of power for any year. The regression equation contains yearly technology parameters to depict the technology available in any given year per unit of cost. A comparison of the structural descriptions with previous computers' structural descriptions shows the list of technology changes.

*Wholistic and holistic approaches to the measurement of technology*

Sahal (1981) suggests two basic ideas of technometrics. In the first approach (*Wholistic*), the state-of-the-art is best specified in terms of a surface of constant probability density given the distribution of technological characteristic. It possible to gauge advances in the surface structure of technical knowledge in terms of generalized distance between the object under consideration. The state-of-the-art at any given point in time is represented by an isodensity contour or probability mountain, rising above the plane. The level of technological capability is given by the height of the mountain. The magnitude of technological change can then be estimated by the difference in the heights of successive mountains. In the second approach (*Holistic*), a technological characteristic is best specified as a vector in an *N*-dimensional space generated by a set of *N* linearly independent elements, such as mass, length, and time. The length of the vector represents the magnitude of a technological characteristic, while the kind of the characteristic is represented by the direction. In this case the state-of-the-art reduces to a point as a consequence of the transmutation of the original parameter space into a dimensional vector space. The successive points at various times constitute a general pattern of technological evolution that evinces a series of *S*-shaped curves. These two approaches are distinct but related.

*Seismic approach to the measurement of technology*

This approach, elaborated by Coccia (2005), categorizes effects of technological change through a scale similar to that used in seismology by Mercalli. In particular, according to the seismic approach of measurement of technological advances, innovations of higher intensity change socioeconomic system with a series of effects on subjects and objects. The impact of the innovation is measured by means of an indicator called Magnitude of Technological Change, which is similar to the magnitude of the Richter scale that measures the energy of earthquakes.

*Technological advances measured with patent data*

These studies are aimed to investigate technological evolution considering the patent data that include information about technologies. Faust (1990) argues that patent indicators allow for a differentiated observation of technological advances before the actual emergence of an innovation, such as technological development in the fields of superconductivity. Wang et al. (2016) investigate technological evolution using US Patent Classification (USPC) reclassification. Results suggest that there exist significant differences among five types of patents based on the USPC reclassification. For instance, patents with Inter-field Mobilized Codes, related to the topics of "Data processing: measuring, calibrating, or testing" and "Optical communications", involved broader technology topics but had a low speed of innovation. Patents with Intra-field Mobilized Codes, mostly in the Computers & Communications and Drugs & Medical fields, tended to have little novelty and a small innovative scope (Wang et al., 2016). Future research in this research field extends the patent sample to subclasses or reclassified secondary USPCs in order to understand the technological evolution within a field in greater detail.

*Technological evolution using a model of technological substitution*

In the context of the measurement of technological advances, Fisher and Pry (1971, p. 75) argue that technological evolution consists of substituting a new technology for the old one, such as the substitution of coal for wood, hydrocarbons for coal, etc. They suggest a simple model of technological substitution that contains only two parameters. Technological advances are represented by competitive substitutions of one method of satisfying a need for another. Fisher and Pry (1971, p. 88) state that: "The speed with which a substitution takes place is not a simple measure of the pace of technical advance . . . . it is, rather a measure of the unbalance in these factors between the competitive elements of the substitution".

Overall, then, although different approaches of measurement of technological advances exist (Sahal, 1981; Arthur and Polak, 2006), a technometrics that measures the evolution of technology considering the inner dynamics of interaction between its subsystems of technology is, at author's knowledge, unknown. To reiterate, this study endeavors to measure and analyses the evolution of technology, as a complex adaptive system, with a new perspective based on coevolution between host technology and its component technologies that has profound effects on long-term development of the whole system of technology.

Next section presents the conceptual framework of this technometrics, which is based on the theory of technological parasitism.

**Conceptual framework for the measurement of the evolution of technology using the theory of technological parasitism**

Evolution of a technology concerns a process of equilibrium governed by the internal dynamics of the technology system (Sahal, 1981, p. 69). In particular, the evolution of a technology can be associated with interaction between a system of technology and its inter-related subsystems. An important step towards the measurement of technological advances is to first clarify the concept of complexity and complex system. Simon (1962, p. 468) states that: "a complex system [is]… one made up of a large number of parts that interact in a nonsimple way …. complexity frequently takes the form of hierarchy, and …. a hierarchic system … is composed of interrelated subsystems, each of the latter being, in turn, hierarchic in structure until we reach some lowest level of elementary subsystem." McNerney et al. (2011, p. 9008) argue that: "The technology can be decomposed into $n$ components, each of which interacts with a cluster of $d-1$ other components" (cf., Gherardi and Rotondo, 2016). Arthur (2009, pp. 18-19) argues that the evolution in technology is due to combinatorial evolution: "Technologies somehow must come into being as fresh combinations of what already exists". This combination of components and assemblies is organized into systems to some human purpose and has a hierarchical and recursive structure. This study here endeavors, starting from concepts just mentioned, to measure technological advances in a framework of host-parasite systems, in a broad analogy with ecology (Coccia and Watts, 2018). Basic concepts of this conceptual framework are:

▫ Technology is defined as a complex system that is composed of more than one component and a relationship that holds between each component and at least one other element in the system. The technology is selected and adapted in the Environment $E$ to satisfy needs and human desires, solve problems in human society and support human control of nature.

▫ Interaction between technologies is a reciprocal adaptation between technologies with interrelationship of information/resources/energy and other physical phenomena to satisfy needs and human wants.

▫ Coevolution of technologies *is* the evolution of reciprocal adaptations in a complex system that generates innovation—i.e., a modification and/or improvement of technologies that interact and adapt in a complex system to expand content of the human life-interests whose increasing realization constitutes progress.

In general, host technologies form a complex system of parts and subsystems that interact in a non-simple way (*e.g.,* batteries and antennas in mobile devices, etc.; cf., Coccia, 2017). In this context, Coccia (2017a) states the theorem of impossible independence of *any* technology that: in the long run, the behavior and evolution of *any* technology is *not* independent from the behavior and evolution of the other technologies. In fact, Sahal (1981, p. 71) argues that: "the evolution of a system is subject to limits only insofar as it remains and isolated system."

Now, suppose that the simplest possible case involves only two technologies, *T1* and *T2,* in a Complex System *S*(T1, T2). The interaction between technologies *T1* and *T2* can be of four typologies to support the evolution of technology:

1. Technological parasitism is a class of relationships between two technologies T1 and T2 in a complex system S where one technology T1 benefits (+) from the interaction with T2, whereas T2 has a negative side (−) from interaction with T1.

2. Technological commensalism is a class of relationships between two technologies where one technology T1 benefits (+) from the other without affecting it (0). The commensal relation is often between a larger host technology and a smaller commensal technology.

3. Technological mutualism is a class of relationships in which each technology benefits from the activity of the other technology (+, +).

4. Technological symbiosis is a class of long-term interactions between two technologies (T1,T2) that evolve together in a complex system S. The symbiotic technologies have a long-run interaction that generates continuous and mutual benefits and, as a consequence, coevolution between a complex system and its interrelated subsystems of technology (++, ++).

The typologies of interaction between technologies that support the evolution of technology can be synthesized in the figure 1.

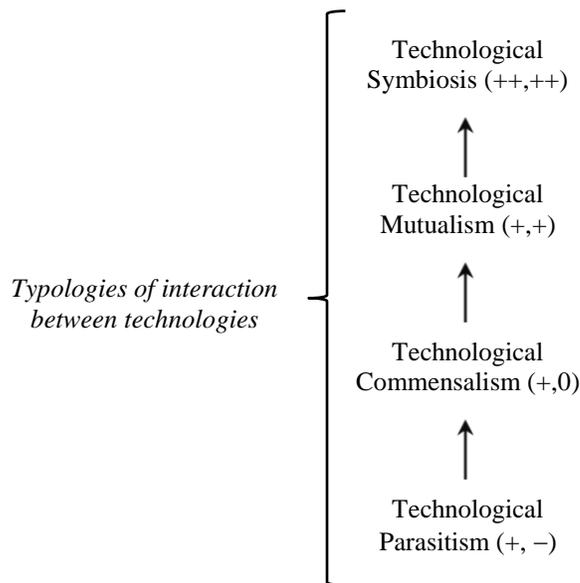

**Figure 1.** Different typologies of interaction between technologies affecting the evolution of the whole system of technology

*Note.* The notions of positive, negative and neutral benefit from interaction between technologies $T_i$ and $T_j$ in S are represented with mathematical symbols +, −, 0 (zero); ++ is a strong positive benefit from long-run mutual symbiotic interaction between technologies $T_i$ and $T_j$ in S (i.e., coevolution of $T_i$ and $T_j$ in S).

In general, parasitism, mutualism, commensalism and symbiosis between technologies do not establish clear cut-offs of these concepts and each relationship represents an end-point of an evolutionary development of technologies in a complex system (cf., Poulin, 2006 for ecological interaction). In particular, parasitism is an interaction that may evolve over time towards commensalism, mutualism and symbiosis to support evolutionary innovations as illustrated in figure 1 (cf., Price, 1991). The symbiosis is also increasingly recognized as an important selective force behind interdependent coevolution of complex systems (cf., Smith, 1991). In general, if technologies increase the grade of interaction within a system, then they support the evolution of the whole system of technology. Moreover, the evolution of technology is a cumulative process that adds new evolutionary phases accompanied by the retention of old one. This process symbolically can be represented in the following manner:

if TS= Technological Symbiosis; TM= Technological Mutualism; TC=Technological Commensalism; TP= Technological Parasitism and $t_1$, $t_2$, $t_3$, $t_4$ are successive periods of time, and *a, b, c, d,* are increasing levels of interaction between technologies, the different types of interaction within a complex technology can be represented:

$TS_{t1}=a$

$TM_{t2}=a+b$

$TC_{t3}=a+b+c$

$TS_{t4}=a+b+c+d$ ∎

Overall, then, the theory of technological parasitism (Coccia and Watts, 2018), shortly describe here, proposes that the interaction between technologies in a complex system tends to generate stepwise coevolutionary processes of a whole system of technology within the "space of the possible" (Wagner and Rosen, 2014, *passim*).

**A proposed technometrics for the evolution of technology in complex systems**

This section endeavors to operationalize the proposed view of the theory of technological parasitism to measure the evolution of technology in the form of a simple model of technological interaction between a host technology H and an interrelated subsystem P. This model focuses on morphological changes in subsystems of technology in relation to proportional changes in the overall host system of technology. This model, based on the principle of allometry, was originally developed by biologists to study the differential growth rates of the parts of a living organism's body in relation to the whole body during the evolution processes (Reeve and Huxley, 1945; Sahal, 1981).

Let *P(t)* be the extent of technological advances of a technology *P* at the time *t* and *H(t)* be the extent of technological advances of a technology H that is a master or host system that interacts with P, at the same time. If both *P* and *H* evolve according to some *S*-shaped pattern of technological growth, such a pattern formally can be represented formally in terms of the differential equation of the logistic

function:

$$\frac{1}{H}\frac{dH}{dt} = \frac{b_1}{K_1}(K_1 - H)$$

We can rewrite the equation as:

$$\frac{K_1}{H}\frac{1}{(K_1 - H)}dH = b_1 dt$$

The integral of this equation is:

$$\log H - \log(K_1 - H) = A + b_1 t$$

$$\log \frac{K_1 - H}{H} = a_1 - b_1 t$$

$$H = \frac{K_1}{1 + \exp(a_1 - b_1 t)}$$

$a_1 = b_1 t$ and $t$ = abscissa of the point of inflection.

The growth of *H* can be described respectively as:

$$\log \frac{K_1 - H}{H} = a_1 - b_1 t \qquad [1]$$

*Mutatis mutandis*, for *P(t)* in similar way of *H(t)*, the equation is:

$$\log \frac{K_2 - P}{P} = a_2 - b_2 t \qquad [2]$$

The logistic curve here is a symmetrical *S*-shaped curve with a point of inflection at 0.5K with $a_1$ = constant depending on the initial conditions, $K_1$ = equilibrium level of growth, and $b_1$ = rate-of-growth parameter.

Solving equations [1] and [2] for *t*, the result is:

$$t = \frac{a_1}{b_1} - \frac{1}{b_1}\log\frac{K_1 - H}{H} = \frac{a_2}{b_2} - \frac{1}{b_2}\log\frac{K_2 - P}{P}$$

The expression generated is:

$$\frac{H}{K_1 - H} = C_1 \left( \frac{P}{K_2 - P} \right)^{\frac{b_1}{b_2}} \qquad [3]$$

When *P* and H are small in comparison with their final value, the following simple model of evolution is obtained:

$$P = A_1 (H)^{B_1} \qquad [4]$$

where $A_1 = \dfrac{K_2}{(K_1)^{\frac{b_2}{b_1}}} C_1$ and $B_1 = \dfrac{b_2}{b_1}$

$B_1$ is the evolutionary coefficient of growth.

The logarithmic form of the equation $P = A_1 (X)^{B_1}$ is a simple linear relationship:

$$\ln P = \ln A_1 + B_1 \ln H \qquad [5]$$

Model of the evolution of technology [5] has linear parameters that are estimated by the Ordinary Least-Squares Method. The value of $B_1$ in the model [5] indicates different patterns of technological evolution and is quantified in real instances in the next section. In particular,

- $B_1 < 1$ whether P evolves at a lower relative rate of change than H; the whole host-system H of technology has a slowed evolution (underdevelopment) over the course of time.

- $B_1$ has a unit value: $B_1 = 1$, then the two technologies P*i* and H*i* have proportional change during their evolution based on a mutualistic interaction (coevolution) between a whole system of technology (H) and its interacting subsystems P*i*. The case of proportional change between elements of a system is also referred to as isometry. In short, the whole system of technology H here has a proportional evolution (growth) over the course of time.

- $B_1 > 1$ whether P*i* evolves at greater relative rate of change than H; this pattern denotes disproportionate technological advance in the structure of a subsystem P as a consequence of change in the overall structure of a technology-host system H. The whole system of technology H has an accelerated evolution (development) over the course of time.

This approach can justify the representational and uniqueness theorem of the measurement in the evolution of technology. Moreover, the results of this model of allometry represented by the coefficient of evolutionary growth can be systematized in an ordinal scale of the evolution of technology (table 2).

Table 2. Scale of the evolution of technology in complex systems

| *Grade of evolution* | *Coefficient of evolutionary growth* | *Type of evolution of technology* | *Evolutionary stages of evolution* | *Predictions* |
|---|---|---|---|---|
| 1 Low | B<1 | Slowed evolution of technology of the whole system | Underdevelopment | Technologies improve slowly over the course of time |
| 2 Average | B=1 | Proportional evolution of technology | Growth | Technologies have a steady-state path of evolution |
| 3 High | B>1 | Accelerated evolution of technology | Development | Technologies are likeliest to evolve rapidly |

Properties of the scale of the evolution of technology (table 2):

a) Technology of higher rank order on the scale has higher technological advances of lower rank order.

b) If a technology has the highest ranking on the scale (i.e., three), it evolves rapidly (development). Vice versa, if a technology has the lowest ranking on the scale (i.e., one), it evolves slowly (underdevelopment).

c) Evolution of technology of higher rank order on the scale has accumulated all previous stages of low rank order of technological evolution and has fruitful mutual symbiotic relationship of growth between a whole system of technology H and its subsystem-components P$i$.

**Materials and methods**

*Data and their sources*

The evolution of technology is illustrated here with realistic examples using historical data of four example technologies: farm tractor, freight locomotive, steam-powered electricity-generating plants and internal-combustion type electric power plants in the USA. Sources of data are tables published by Sahal (1981, pp.319-350, originally sourced from trade literature; cf., also Coccia, 2018). Note that data from the earliest years and also the war years are sparse for some technology.

*Measures*

Technological parameters that measure the evolution of technology are given by Functional Measures of Technology (FMT) over the course of time to take into account both major and minor innovations (Sahal, 1981, pp. 27-29).

Measures of technological parameters for farm tractor are:

− fuel-consumption efficiency in horsepower-hours over 1920-1968 CE indicates the technological advances of engines (a subsystem) of farm tractors. This FMT represents the dependent variable P of the model [5].

− mechanical efficiency (ratio of drawbar horsepower to belt or power take-off –PTO- horsepower) over 1920-1968 CE indicates the technological advances of farm tractor. This FMT represents the explanatory variable H of the model [5].

For freight locomotive, FMTs are:

− Tractive efforts in pound over 1904-1932 CE indicates the technological advances of locomotive in the USA. This FMT represents the dependent variable P of the model [5].

− Total railroad mileage over 1904-1932 CE indicates the technological advances of the infrastructure system of railroad in the USA. This FMT represents the explanatory variable of the model [5].

For steam-powered electricity-generating technology, FMTs are:

− Average fuel-consumption efficiency in kilowatt-hours per pound of coal over 1920-1970 CE indicates the technological advances of boiler, turbines and electrical generator. This FMT represents the dependent variable P of the model [5].

− Average scale of plant utilization (the ratio of net production of steam-powered electrical energy in millions of kilowatt-hours to number of steam powered plants) over 1920-1970 CE indicates the technological advances of the overall electricity-generating plants. This FMT represents the explanatory variable of the model [5].

For internal-combustion type electric power technology, FMTs are:

− Average fuel-consumption efficiency in kilowatt-hours per cubic foot of gas 1920-1970 CE indicates the technological advances of boiler, turbines and electrical generator. This FMT represents the dependent variable P of the model [5].

− Average scale of plant utilization (the ratio of net production of electrical energy by internal-combustion type plants in millions of kilowatt-hours to total number of these plants) over 1920-1970 CE indicates the technological advances of the overall electricity-generating plants. This FMT represents the explanatory variable of the model [5].

*Model and data analysis procedure*
Model [5] of the technological evolution implemented in real instances here is:

$Ln\ P_t = ln\ a + B\ Ln\ H_t + u_t$     (with $u_t$ = error term)          [6]

***a*** is a constant

$P_t$ will be the extent of technological advances of technology that represents a subsystem of the Host technology $H$ at time $t$

$H_t$ will be the extent of technological advances of technology at time $t$ that represents the host technology of a interacting subsystem $P$; it is the driving complex system of overall interrelated subsystems of technology.

*Remark*: Let $b_1$ and $b_2$ be the growth rates of *P* and *H* respectively, such that $B = \dfrac{b_1}{b_2}$ measures the relative growth of *P* in relation to the growth of H. *B*<1, ≥ 1

Data are transformed in logarithm to obtain normal distributions that are appropriate to estimate the relationships [6] with regression analysis using Ordinary Least Squares method. Statistical analyses are performed with the Statistics Software SPSS® version 24.

**Case studies of the evolution of technology in the agriculture, rail transport and electricity generation**

The evolution of technology modelled with an approach of allometry [6] is illustrated here with realistic examples using historical data of farm tractor, freight locomotive, steam-powered electricity-generating and internal-combustion type electric power technology in the USA.

Table 3 shows the descriptive statistics.

*Table 3* – Descriptive statistics

|  | LN Fuel consumption efficiency in horsepower hours (Engine of Tractor) | LN Mechanical efficiency ratio of drawbar horsepower to belt (Tractor) | LN Tractive efforts in pound (Locomotive power) | LN Total railroad mileage (Infrastructure for Locomotive) |
|---|---|---|---|---|
| N | 44 | 44 | 29 | 29 |
| Mean | 2.13 | 4.19 | 10.43 | 12.86 |
| Std. Deviation | 0.27 | 0.146 | 0.22 | 0.11 |
| Skewness | -0.76 | -0.68 | -0.21 | -1.04 |
| Kurtosis | -0.83 | -0.56 | -1.19 | -0.06 |

|  | LN Average fuel consumption efficiency in kwh per pound of coal (turbine and various equipment in steam-powered plant) | LN Average scale of steam-powered Plants (steam-powered plant) | LN Average fuel consumption efficiency in kwh per cubic food of gas (turbine and various equipment in internal-combustion plant) | LN Average scale of internal-combustion plants (internal-combustion plant) |
|---|---|---|---|---|
| N | 51 | 51 | 51 | 51 |
| Mean | -0.25 | 4.85 | -2.75 | 0.51 |
| Std. Deviation | 0.34 | 1.43 | 0.33 | 0.85 |
| Skewness | -0.67 | -0.17 | -0.67 | 0.02 |
| Kurtosis | -0.09 | -1.26 | 0.04 | -1.64 |

*Results of the evolution of farm tractor technology (1920-1968)*

Table 4 shows that $B = 1.74$, i.e., $B > 1$, the subsystem component technology of engine has a disproportionate growth in comparison with overall farm tractor. This indicates a high grade of evolution in the scale of the evolution of technology that represents a development of the whole system of farm tractor (cf. Figure 2).

*Table 4* – Estimated relationship for farm tractor technology based on allometric equation

*Dependent variable*: LN fuel consumption efficiency in horsepower hours (technological advances of engine for tractor –sub-system component technology) at $t = 1920, \ldots, 1968$

|  | Constant $\alpha$ (St. Err.) | Evolutionary coefficient $\beta=B$ (St. Err.) | $R^2$ adj. (St. Err. of the Estimate) | F (sign.) |
|---|---|---|---|---|
| Farm tractor | −5.14*** | 1.74*** | 0.85 | 256.44 |
|  | (0.45) | (0.11) | (0.10) | (0.001) |

*Note*: ***Coefficient $\beta$ is significant at 1‰; Explanatory variable is LN mechanical efficiency ratio of drawbar horsepower to belt (technological advances of farm tractor –Host technology), $t = (1920–1968)$

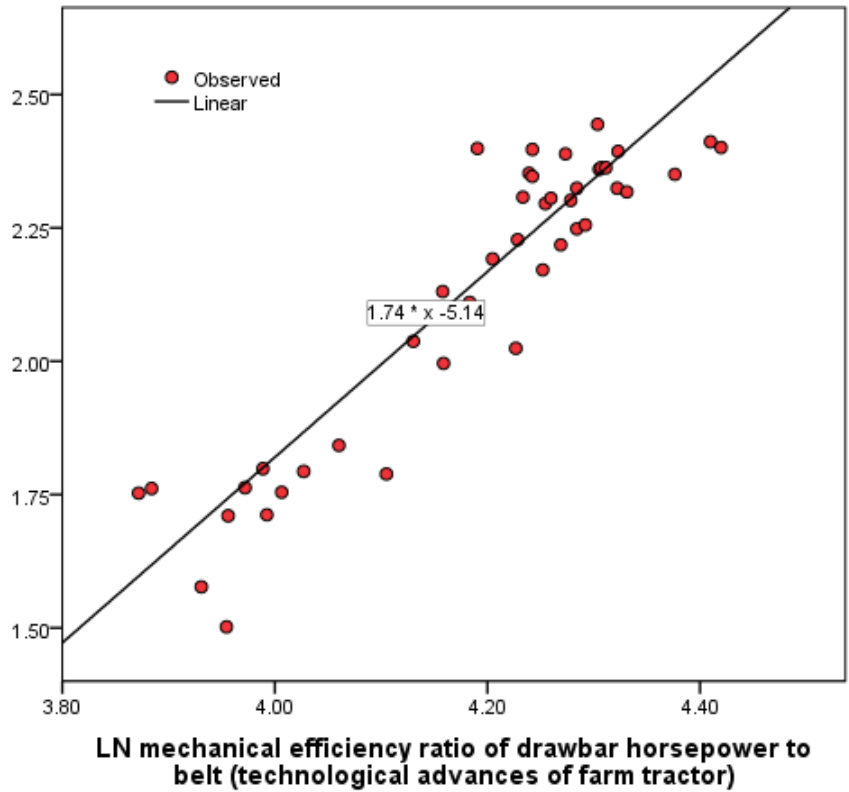

**Figure 2.** Trend and estimated relationship of the evolution of farm tractor technology

This high evolution of farm tractor technology is due to numerous advances and radical innovations over time, such as the diesel-powered track-type tractor in 1931, low-pressure rubber tires in 1934 and introduction of remote control in 1947 that made possible improved control of large drawn implements.

The development of the continuous running power takeoff (PTO) also in 1947 allowed the tractor's clutch to be disengaged without impeding power to the implements. In 1950 it is introduced the 1000-rpm PTO for transmission of higher power, whereas in 1953 power steering was applied in new generations of tractor. Moreover the PTO horsepower of the tractor has more than doubled from about 27hp to 69hp over 1948-1968; finally, dual rear wheels in 1965 and auxiliary front-wheel drive and four-wheel drive in 1967 (Sahal, 1981). These radical and incremental innovations have supported the accelerate evolution of the farm tractor technology over time.

*Results of the evolution of freight locomotive technology (1904–1932)*

Table 5 shows that $B = 1.89$, i.e., $B > 1$ indicates a stage of development of freight locomotive in the complex system of rail transportation technology (see, Figure 3).

*Table 5* – Estimated relationship for freight locomotive technology based on allometric equation

*Dependent variable*: LN Tractive efforts in pound of locomotive (technological advances), $t = (1904–1932)$

|  | Constant $\alpha$ (St. Err.) | Evolutionary coefficient $\beta=B$ (St. Err.) | $R^2$ adj. (St. Err. of the Estimate) | F (sign.) |
|---|---|---|---|---|
| Farm tractor | −13.87*** | 1.89*** | 0.91 | 270.15 |
|  | (1.48) | (0.12) | (0.07) | (0.001) |

*Note*: ***Coefficient β is significant at 1‰; Explanatory variable is LN Total railroad mileage (technological advances of the infrastructure –Host technology) at $t =1904, …, 1932$

This development of freight locomotive can be explained with a number of improvements, such as in 1906 it is introduced the compound engine that improved the tractive effort. In 1912 the first mechanical stoker to use the steam-jet overfeed system of coal distribution is used and another technological advance in this area was the substitution of pneumatically operated power reverse gear for the hand lever. In 1916 it is introduced the unit drawbar and radial buffer that eliminated the need for a safety chain in coupling the engine and tender together. Further advances are due to the adoption of cast-steel frames integral with the cylinder, the chemical treatment of the locomotive boiler water supply and the introduction of roller bearings over 1930s. In particular, these technical developments reduced the frequency of maintenance work. Subsequently, the continuous modification of the steam locomotive with reciprocating engine has led to diesel-electric locomotive by the mid-1940s (Sahal, 1981). These technological developments supported the accelerated evolution of this technology.

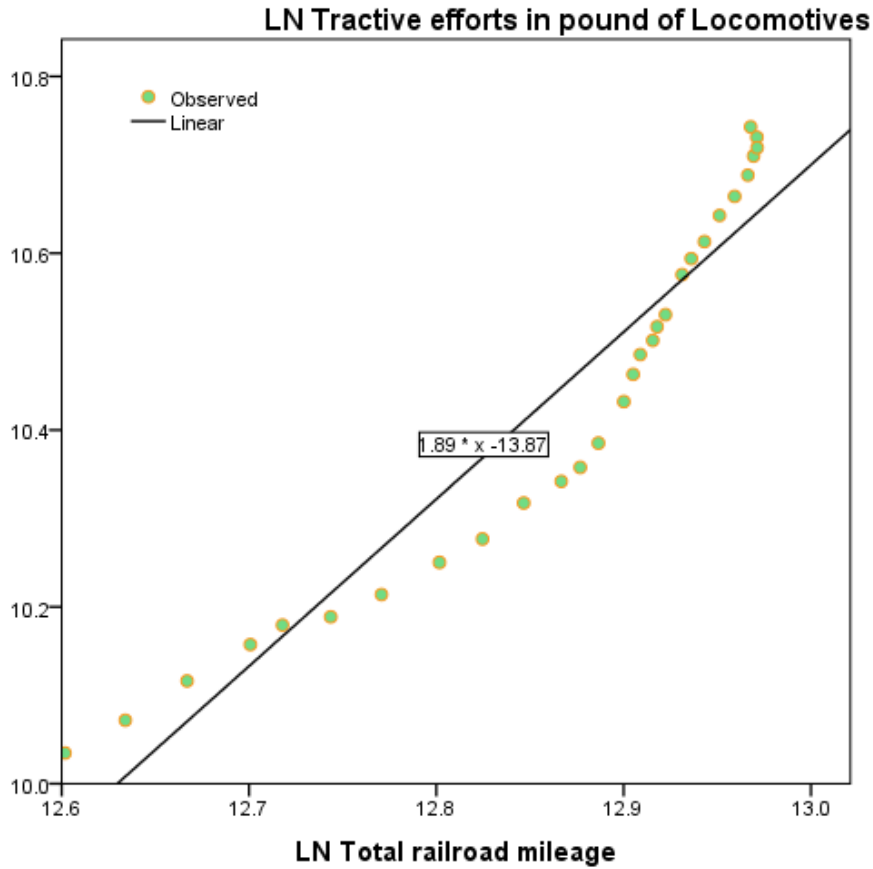

**Figure 3.** Trend and estimated relationship of the evolution of freight locomotive technology

*Results of the evolution of electricity generation technology (1920-1970)*

Table 6 shows that $B = 0.23$, i.e., $B < 1$ (see also, Figure 4).

*Table 6 – Estimated relationship for the steam-powered electricity-generating technology (1920-1970)*

| *Dependent variable*: LN Average fuel consumption efficiency in kwh per pound of coal (technological advances of turbine and various equipment) | | | | |
|---|---|---|---|---|
| | *Constant* $\alpha$ *(St. Err.)* | *Evolutionary Coefficient* $\beta=B$ *(St. Err.)* | $R^2$ *adj. (St. Err. of the Estimate)* | *F (sign.)* |
| Turbine for steam-powered electricity | −1.35*** (0.04) | 0.23*** (0.01) | 0.93 (0.09) | 675.12 (0.001) |

*Note*: ***Coefficient β is significant at 1‰; Explanatory variable is Average scale of steam-powered plants (Host technology) at $t = 1920, \ldots, 1970$

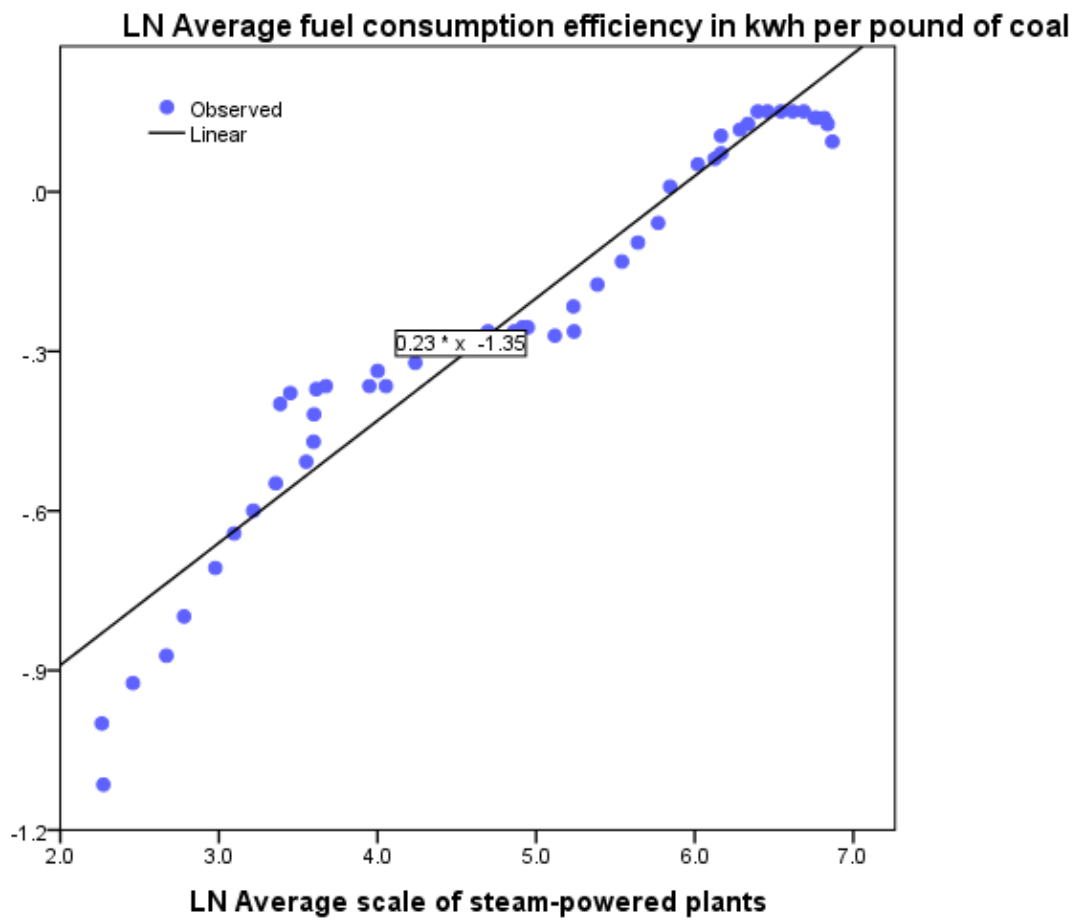

**Figure 4.** Trend and estimated relationship of the evolution steam-powered electricity-generating plants (1920-1970)

Table 7 shows for internal-combustion type electric power plants similar results to steam-powered plants: $B = 0.35$, i.e., $B < 1$. In short, evolution of technology in the generation of electricity is low and driven by an evolutionary route of underdevelopment over the course of time (see, Figure 4 and 5).

This evolution of technology in the generation of electricity is associated with available natural resources, the increase in steam pressure and temperature made possible by advances in metallurgy, the use of double reheat units and improvements in the integrated system man-machine interactions to optimize the operation of plants. In general, the rate of technological evolution in the electricity generation plants has slowed down because of: "the deterioration in the quality of fuel and of constraints imposed by environmental conditions….other main reasons: First, increased steam temperature requires the use of more costly alloys, which in turn entail maintenance problems of their own…. Thus there has been a decrease in the maximum throttle temperature from 1200 °F in 1962, to about 1000 °F in 1970. Second, there has been lack of motivation to increase the efficiency in the use of gas in both steam-powered and internal-combustion plants because of the artificially low price of fuel due to Federal Power Commission's wellhead gas price regulation. Finally, … there has been a slowdown in generation efficiency due to heavy use of low-efficiency gas turbines necessitated by delays in the construction of nuclear power plant" (Sahal, 1981, p. 184).

*Table 7* – Estimated relationship for internal-combustion type electric power plants (1920-1970)

*Dependent variable*: LN Average fuel consumption efficiency in kwh per cubic food of gas (technological advances of turbine and various equipment)

|  | Constant $\alpha$ (St. Err.) | Evolutionary coefficient $\beta=B$ (St. Err.) | $R^2$ adj. (St. Err. of the Estimate) | F (sign.) |
|---|---|---|---|---|
| Turbine for electric energy by Internal-combustion | −2.93*** (0.02) | 0.35*** (0.02) | 0.81 (0.14) | 213.63 (0.001) |

*Note*: ***Coefficient β is significant at 1‰; Explanatory variable is LN Average scale of internal-combustion plants (Host technology) at $t = 1920, …, 1970$

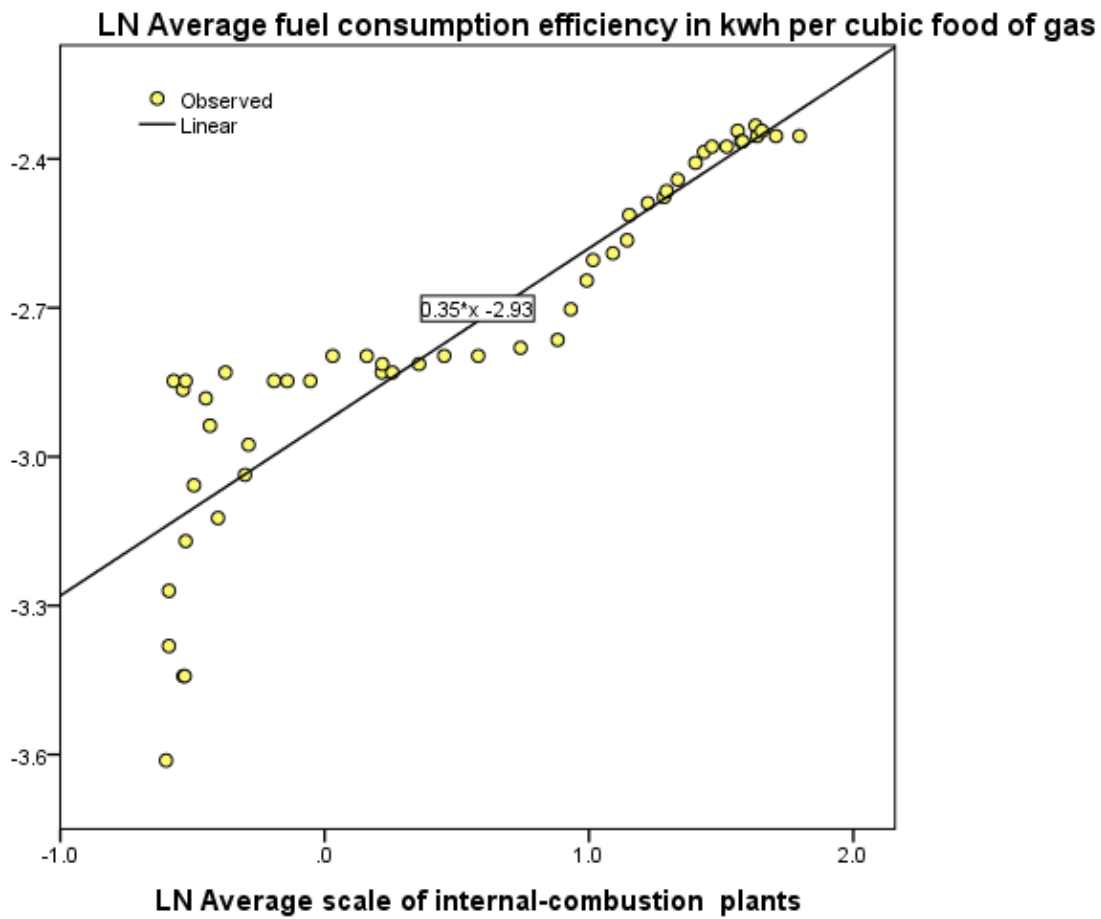

**Figure 5.** Trend and estimated relationship of the evolution of internal-combustion technology for electric power plants (1920-1970)

**Discussion**

The evolution of technology needs a unifying perspective to support the comprehensive measurement of technological advances and aid in its comprehension. This article proposes a new perspective for the evolution of technology that is adapted from ecology and is modelled with a simple model of morphological change to measure and explain the technological development of the whole system of technology over the long run. As a matter of fact, some scholars argue that technologies and technological change display numerous life-like features, suggesting a deep connection with biological evolution (Basalla, 1988; Erwin and Krakauer, 2004; Solé et al., 2011; Wagner and Rosen, 2014). In general, Darwinian evolution and ecology seem to support possible explanations of technology evolution as it has done for species development (Basalla, 1988). This study extends the broad analogy between technological and biological evolution to more specifically focus on the potential of a technometrics based on interaction between technologies in complex systems, but fully acknowledge that interaction between technologies is not a perfect analogy of biological/ecological interaction; of course, there are differences (Ziman, 2000; Jacob, 1977; Solé *et al.,* 2013). For studying technical change, though, the analogy with biology and ecology is a source of inspiration and ideas because it has been studied in such depth and provides a logical structure of scientific inquiry in these research fields. The study here seems to provide an appropriate theoretical framework to measure evolution of technology and explain possible evolutionary pathways of complex systems of technology. The evolution of technology here is based upon a simple assumption that technologies are complex adaptive systems that interact in a nonsimple way with other technologies and its interrelated subsystems of technology. A consequence of this approach is that the evolution of technology is a multidimensional process of interaction within and between technologies, such that technological advances of a system can slow down when it remains an isolated system and does not interact with other technologies.

Moreover, the inner dynamics of the evolution of technology is based on S-shaped growth curve of technological advances both for the whole system of technology and for its interrelated subsystem components. This approach is formalized with a simple model that contains only two parameters and provides the coefficient of evolutionary growth, which is useful to measure the typology of evolution of technology and which technologies are likeliest to evolve rapidly. In particular, the technometrics here provides three simple grades of the evolution of technology according to the coefficient of evolutionary growth B: underdevelopment (B<1), growth (B=1) and development of the whole system of technology (B>1). The theory is illustrated with realistic examples that show different patterns of evolution of technologies.

The quantification of the coefficient of evolutionary growth, called B, can suggest reliable predictions of the long-term development of technology. In particular,

1. Evolution of technology in the form of development of the whole system is governed by a process of disproportionate growth in its subsystems (B>1) as a consequence of change in the overall system of the host technology (e.g., technological development of farm tractor and freight locomotive).

2. Evolution of technology is slowdown when component subsystems of a technology have low changes as a consequence of the whole system (B<1), generating underdevelopment of the whole system of technology over the course of time (e.g., technology in the electricity generation).

3. The long-run evolution of a technology depends on the behavior and evolution of associated technologies. To put it differently, long-run evolution of a specific technology is enhanced by the integration of two or more technologies that generate co-evolution of system innovations.

4. Technologies having a symbiotic growth with other associated technologies (B>1) advance rapidly, whereas technologies with low growth of its parts (B<1) improve slowly.

5. Isolated system of technology with low interaction between systems and among the parts of those systems is subject to limits of growth.

Sahal (1981, p. 69) argues that the dynamics of a system is affected by its history and associated evolutionary processes tend to be self-generating and also self-constraining of its growth. The evolution of technology modelled here is due to processes of learning driven by interaction between different technological devices and its subsystems that determines the scope for the utilization of a technology, technological guideposts and innovation avenues (cf., Sahal, 1981; Nelson and Winter, 1982). Sahal (1981, p. 82, original italics) also argues that: "the role of learning in the *evolution* of a technique has profound implications for its *diffusion* as well". In particular, findings show that the evolution of technologies is affected by scientific and technological advances (e.g., for farm tractor and freight locomotive) but it is also affected by socio-institutional environments that can slowdown the technological progress (e.g., technological advances for efficiency of powered plant in the generation of electricity).

This finding could aid technology policy and management of technology to design best practices for supporting development of new technology, and as a consequence, industrial and economic change, and general technological progress of human society.

Proposed theory here has also a number of implications for the analysis of nature, source and evolution of technology. Some of the most obvious implications, without pretending to be comprehensive are as follows. This study contributes to the literature on evolution of technology by detailing the importance of specific stages during the evolutionary patterns of technological innovation based on interaction between technologies in complex systems. The dynamic characteristic underlying the proposed technometrics here may also help better understand the linkages between technologies, such as how technological parasitism, commensalism, mutualism and symbiosis drive the evolutionary pathways of complex systems of technology and technological diversification over time and space. This theory here also extends the literature on technological evolution identifying some important but overlooked process governed by the dynamics of technology based on interaction between technology and its interacting subsystems. In fact, the approach detects phases and typologies that guide the evolution of technology in complex systems. The suggested approach of technometrics here is consistent with the well-established literature by Arthur (2009) as well as with studies that consider structural innovations and systems innovations based on integration of two or more symbiotic technologies (Sahal, 1981).

The theoretical framework here can suggest some general properties of the evolution of technology as a complex system. In particular, evolution of technology generally involves underlying processes of interaction between technology and its subsystems given by: (1) disproportionate growth of its subsystems and (2) increase in the complexity of the structure of technology over time.

However, drawbacks of the approach here to measurement of the evolution of technology are that useful data was not available in sufficient quality. Future efforts in this research field require a substantial amount of data of technological parameters that measure several characteristics of the dynamics of the whole system of technology and component subsystems to provide more empirical evidence of the evolution of technologies and new technological trajectories.

To conclude, the proposed technometrics here based on the ecology-like interaction between technologies—may lay the foundation for development of more sophisticated concepts and theoretical frameworks in technological studies. In particular, this study constitutes an initial significant step in measuring evolution of technology considering the interaction between technologies in complex systems. However, identifying comprehensive technometrics in the domain of different complex systems of technology, affected by manifold factors, is a non-trivial exercise. Wright (1997, p. 1562) properly claims that: "In the world of technological change, bounded rationality is the rule."

**Appendix: simple indices for measuring evolution and coevolution of technology**

☐ *Case of 2 technologies*

Assumption: T*i*=technology *i*, with *i*=1, 2

$T_1$= Technology 1, $T_2$= Technology 2

$T_1$ and $T_2$ have a close technological interaction with $T_1$=host (or master) technology and $T_2$=parasitic technology.

Suppose:

$t_1$= life cycle period of technology $T_1$

$t_2$= life cycle period of technology $T_2$

$c_1$=number of change of generations of technology $T_1$ in $t_1$

$c_2$=number of change of generations of technology $T_2$ in $t_2$

Evolution Technology 1 : $Ev_1 = \dfrac{c_1}{t_1}$ [1A]

Evolution Technology 2 : $Ev_2 = \dfrac{c_2}{t_2}$ [2A]

$CV_i$ is the index of technological coevolution given by:

$CV_{i=1,2} = Ev_1 \times Ev_2$ [3A]

$CV_i \in [0, +\infty]$ $with$ $CV_i > 0.1$ ( $coevolution$), $MinCV_i = 0$ ($no\ coevolution$) See Figure 1A

- *Generalization*

Assumption: T$i$=Technology $i$, $i=1,2, …, n$

With $T_1$=host (or master) technology; $T_{2, …, n}$ =parasitic technologies

$T_i$ with $i=1, …, n$ has a close technological interaction in a system of technology

$Ev_1$= Evolution Technology $T_1 = \dfrac{c_1}{t_1}$

$Ev_i$= Evolution Technology $T_2 = \dfrac{c_i}{t_i}$

$Ev_n$= Evolution Technology $T_n = \dfrac{c_n}{t_n}$

$CVT_i$ is the general index of technological coevolution given by:

$$CVT_{i=1,2,…,n} = Ev_1 \times Ev_2 \times … Ev_i \times … Ev_n \qquad [4A]$$

*Property* of *CVTi*: the long-run rate of coevolution of the whole system of technologies is higher than the rate of evolution of subsystems of technologies:

$$CVT_{i=1,2,…,n} \geq Ev_1, Ev_2, …, Ev_i, … Ev_n \qquad [5A]$$

*Proof*: This inequality [5A] is due to the characteristics of hierarchic and nearly decomposable systems, *sensu* Simon (1962).

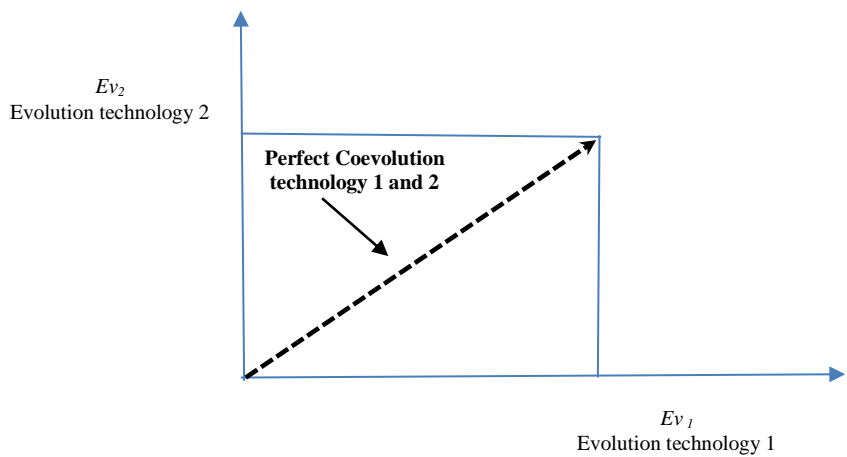

**Figure 1A.** Coevolution of technology 1 and 2 in the 2D space. The 45° line indicates the perfect coevolution of technologies 1 and 2: same rates *Ev* of evolutionary change over time. *Note*: in case of *n* technologies the coevolution is in a multidimensional *N*-space (Coccia, 2017).

*Application* (Coccia, 2017).

The specific host-parasitic system of *i*Phone technology- host (or master) - and WhatsApp - parasitic technology is an interesting case study.

The first release of *i*Phone was June 29, 2007 with the 1th generation. In September 16, 2016, after 9 years, is released the 10th generation.

Rate of evolutionary change *Ev* [1] of *i*Phone technology is given by:

$$Ev_1 = (iPhone) = \frac{10}{9} = 1.11$$

The parasitic technology WhatsApp 2.0 was first released in App stores in 2009. The product has had 14 new improvements of products until the 2016.

The rate of evolutionary change *Ev* [1] of WhatsApp technology is:

$$Ev_2 = (WhatsApp) = \frac{14}{7} = 2.0$$

Hence, this specific case shows that parasitic technology tends to have, in average, a higher evolutionary change than host (or master) technology.

The technological coevolution of host (or master) complex systems with subsystems of parasitic technologies can be calculated with the Equation [3A]. The case study here shows that $CV_i$ with $i$=1, 2 (where 1=parasitic technology, WhatsApp; 2=host (or master) technology, *i*Phone) is:

$$CV_{i=1,2} = Ev_1 \times Ev_2 = 1.11 \times 2.0 = 2.22$$

$$CV_{i=1,2} \geq Ev_1, Ev_2$$